# Estado del clima en Jalisco: temporada de lluvias y comportamiento extremo de la temperatura en 2024

Climate Status in Jalisco: Rainy Season 2024 and Extreme Temperature Behavior in 2024

Julio Eduardo Zamora-Salvador[1]
ORCID: https://orcid.org/0009-0002-0425-000X
Alma Delia Ortíz-Bañuelos[1]
ORCID: https://orcid.org/0000-0001-6483-9102
Stephany Paulina Arellano-Ramírez[1-2]
ORCID: https://orcid.org/0000-0002-7963-5505
Carlos Román-Castañeda[1]
ORCID: https://orcid.org/0009-0001-5271-0851
Armando González-Figueroa[1]
ORCID: https://orcid.org/0009-0000-5188-9215
Hector Hugo Ulloa-Godínez[1]
ORCID: https://orcid.org/0000-0002-1888-628X
Mario E. García Guadalupe[1]
ORCID: https://orcid.org/0000-0003-2179-4818
Mauricio López-Reyes[1-3-4*]
ORCID: https://orcid.org/0000-0001-8138-4269

[1]Universidad de Guadalajara. Departamento de Física del Centro Universitario de Ciencias Exactas e Ingenierías. Instituto de Astronomía y Meteorología (IAM). Guadalajara, Jalisco, México.
[2]Universidad de Guadalajara. Centro Universitario de Ciencias Sociales y Humanidades. Centro de Estudios Estratégicos para el Desarrollo (CEED). Guadalajara, Jalisco, México.
[3]Universidad Complutense de Madrid (UCM). Departamento de Física de la Tierra y Astrofísica. Madrid, España.
[4]Departamento de Investigación. Instituto Frontera. Tijuana, México.
*Autor para correspondencia:
mauricio.lopez@academicos.udg.mx

**Resumen**

Este artículo analiza las características de la temporada de lluvias y las temperaturas extremas registradas en el estado de Jalisco, México, durante 2024. A partir de datos de reanálisis ERA5, CHIRPS y observaciones de trece estaciones meteorológicas, se identificaron patrones de anomalías en la precipitación y la temperatura. Los resultados muestran un inicio tardío y una conclusión anticipada de la temporada de lluvias, con anomalías positivas en las regiones centro y sur, y déficits significativos en la franja costera. Además, 2024 destacó como el año más cálido registrado, con una anomalía promedio de 2.3°C en el periodo mayo–octubre, superando el récord de 2023. Este calentamiento probablemente es atribuido a la combinación del cambio climático antropogénico y un evento de El Niño, alterando los patrones atmosféricos locales y afectando la distribución de las lluvias. Los hallazgos evidencian la vulnerabilidad climática de Jalisco, especialmente en el sector agropecuario, que depende de patrones climáticos estables. Este estudio subraya la necesidad de fortalecer las redes de monitoreo, implementar modelos predictivos y diseñar estrategias de adaptación para mitigar los impactos en sectores clave como la agricultura y los recursos hídricos.

**Palabras clave:** Precipitación, clima, Jalisco, variabilidad climática.

**Abstract**

This article analyzes the characteristics of the rainy season and extreme temperatures recorded in the state of Jalisco, Mexico, during 2024. Using ERA5 reanalysis data, the CHIRPS product, and observations from thirteen weather stations, precipitation and temperature anomaly patterns were identified. The results show a late onset and an early conclusion of the rainy season, with positive anomalies in the central and southern regions and significant deficits along the coastal strip. Furthermore, 2024 was the warmest year on record, with an average anomaly of 2.3°C during May–October, surpassing the 2023 record. This warming probably is attributed to the combination of anthropogenic climate change and an El Niño event, altering local atmospheric patterns and affecting rainfall distribution. The findings highlight Jalisco's climate vulnerability, especially in the agricultural sector, which depends on stable climatic patterns. This study underscores the need to strengthen monitoring networks, implement predictive models, and design adaptation strategies to mitigate impacts on key sectors such as agriculture and water resources.

## Introducción

En el estado de Jalisco, México, las precipitaciones más significativas ocurren entre los meses de junio y octubre, durante la denominada temporada de lluvias, la cual coincide con la temporada de ciclones tropicales en el Pacífico oriental, oficialmente definida del 15 de mayo al 30 de noviembre (NOAA, 2013). Aunque estas fechas proporcionan un marco general, la ocurrencia de lluvias en la región presenta una alta variabilidad interanual, influenciada por fenómenos atmosféricos y océano-atmósfera como la Oscilación del Sur-El Niño (ENOS) y la Oscilación Madden-Julian (MJO), así como por la ubicación de la Zona de Convergencia Intertropical (Pérez-Dolores, 2023; Estrada-Porrúa *et al*., 2015; Ugalde *et al*., 2024).

La fase cálida de ENOS, conocida como El Niño, tiende a reducir las precipitaciones estivales en Jalisco debido al debilitamiento de la convección y al fortalecimiento de un anticiclón subtropical que altera los patrones de circulación atmosférica (Pavía *et al*., 2006; Bravo *et al*., 2010). Por otro lado, durante La Niña, las lluvias en la región suelen intensificarse, favoreciendo actividades agrícolas dependientes del temporal. Esta dualidad refleja la sensibilidad del clima de Jalisco a los cambios en los patrones globales, lo que enfatiza la necesidad de estudiar detalladamente las anomalías climáticas locales.

La interacción entre la variabilidad climática y el sector agropecuario de Jalisco es particularmente crítica, ya que la agricultura de temporal, que representa una proporción significativa de la actividad agrícola estatal, depende casi exclusivamente de las precipitaciones estacionales. Cambios en los patrones de lluvia, ya sean déficits o excesos, afectan directamente los rendimientos de cultivos clave como el maíz y diferentes variedades de agave, especialmente el *Agave tequilana* Weber var. azul, este último fundamental para la industria del tequila, uno de los pilares económicos de la región (Rodríguez-Moreno *et al*., 2014; Magaña *et al*., 2003). Adicionalmente, las temperaturas extremas registradas en los últimos años, exacerbadas por el cambio climático, incrementan el estrés térmico en cultivos y ganado, afectando la productividad y calidad (Lobell *et al*., 2011; Anderson *et al*., 2019; Quiroga *et al*., 2020). Estas condiciones climáticas adversas subrayan la necesidad de estrategias de adaptación basadas en una comprensión más profunda de las dinámicas atmosféricas locales y su impacto en los sistemas agropecuarios.

En este contexto, la variabilidad en las precipitaciones tiene implicaciones socioeconómicas significativas. Mientras los agricultores dependen de una temporada lluviosa regular para maximizar la producción de cultivos, sectores como el comercio informal y las actividades urbanas enfrentan desafíos asociados a lluvias intensas y prolongadas. Este artículo busca analizar el comportamiento de la temporada de lluvias en 2024, evaluando sus características mensuales y estacionales en comparación con el clima presente. Además, se exploran las temperaturas extremas registradas, destacando que 2023 y 2024 han sido los años más cálidos desde que se tienen registros, atribuible tanto al cambio climático antropogénico como al fenómeno de El Niño (WMO, 2023; NOAA, 2024).

Para alcanzar estos objetivos, se emplearon datos de estaciones meteorológicas, reanálisis ERA5 y el producto CHIRPS (Climate Hazards Group InfraRed Precipitation with Station Data), los cuales fueron procesados e interpolados para generar una representación espacial y temporal detallada. Los resultados permitirán comprender mejor la dinámica climática de la región y sus implicaciones en sectores clave como la agricultura y la gestión de recursos hídricos.

Este artículo se organiza de la siguiente manera: en la sección 2, se describen los datos y la metodología empleada; en la sección 3, se presentan los resultados relacionados con el comportamiento de la temporada de lluvias y las temperaturas, incluyendo el análisis de un caso específico de tormenta que afectó al Área Metropolitana de Guadalajara (AMG). Finalmente, la sección 4 está dedicada al resumen y los comentarios finales.

## Materiales y Método
### Zona de estudio

La zona de estudio está ubicada dentro de la zona tropical norte del planeta, con coordenadas extremas de -106° O y -101° O de longitud, así como 18°N y 23° N de latitud (Figura 1), este territorio tiene un relieve compuesto de llanuras, sierras y zonas volcánicas, destacando el Eje Neovolcánico Transversal (Ferrari *et al*., 2012); 118, 116.6 hectáreas de su territorio está compuesto por cuerpos de agua, como ríos, presas, lagunas y un lago (INEGI, 2013); otra característica de la región, es que cuenta con dos estaciones muy marcadas: la temporada de lluvias y la temporada seca el resto del año (Magaña *et al.*, 2003).

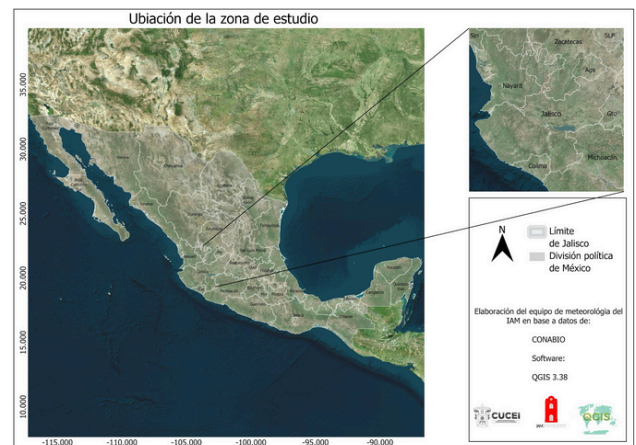

Figura. 1. Región de estudio, Estado de Jalisco, México.
Fuente: Elaboración propia.

**Datos y métodos**

Para este estudio, se utilizaron datos diarios de precipitación del producto CHIRPS con una resolución espacial de 0.25°, abarcando el periodo de mayo a octubre de 2024. La climatología de precipitación se calculó con los datos del periodo base 1991-2020, y a partir de esta se determinaron las anomalías mensuales de precipitación para 2024.

Para las anomalías de temperatura a 2 metros en 2024, se emplearon datos horarios del reanálisis ERA5 con una resolución espacial de 0.25° (Hersbach *et al.*, 2020). La climatología de temperatura también se construyó utilizando ERA5 para el periodo 1991-2020. Para destacar el evento extremo de temperaturas en Jalisco durante 2024, se utilizaron los datos de temperatura a 2 metros en el periodo climático (1981-2020) para calcular los promedio mensuales en Jalisco. A partir de estos promedios, se generó una distribución de densidad de probabilidad (FDP) utilizando un kernel de estimación de densidad (KED) con la biblioteca Seaborn en Python. Con el propósito de caracterizar la situación meteorológica dominante durante la temporada más cálida del año en Jalisco y que corresponde con el inicio de la temporada de lluvias, se calcularon anomalías de altura geopotencial en 500 hPa ($Z_{500}$) y humedad específica en 850 hPa ($q_{850}$) correspondientes a los meses de mayo y junio a partir de datos de ERA5.

Cuadro 1. Información sobre la ubicación de las estaciones meteorológicas utilizadas.

| Nombre de la estación | Municipio | Latitud | Longitud | Altitud |
|---|---|---|---|---|
| GEOSMET_Matatlán | Zapotlanejo | 20.62° N | -103.06° O | 1,550 m |
| GEOSMET_GDL_sur | Zapopan | 20.72° N | -103.38° O | 1,563 m |
| Mirador de San Ysidro | Zapopan | 20.61° N | -103.41° O | 1,652 m |
| Preparatoria UdeG No. 16 | Tlaquepaque | 20.64° N | -103.31° O | 1,550 m |
| GEOSMET_Arandas | Arandas | 20.70° N | -102.35° O | 2,087 m |
| GEOSMET_Agrometrika | Atotonilco | 20.59° N | -102.50° O | 1,920 m |
| MeteoCiudadGuzman | Zapotlán el Grande | 19.90° N | -103.46° O | 1,504 m |
| UEPCBJ-Jalocote | Autlán de Navarro | 19.77° N | -104.36° O | 924 m |
| La Cuesta Jal | Talpa de Allende | 20.38° N | -104.82° O | 1,320 m |
| Estrella del Mar | Puerto Vallarta | 20.65° N | -105.22° O | 5 m |
| UEPCBJ-San Sebastián | San Sebastián | 20.76°N | -104.86°O | 1,480 m |
| UEPCBJ-Tequila | Tequila | 20.88°N | -103.83°O | 1,206 m |
| San José de Tepozán | Lagos de Moreno | 21.35°N | -101.92°O | 1,945 m |

Con el fin de mejorar la representación espacial y temporal de las variables estudiadas, los datos de CHIRPS y ERA5 fueron forzados utilizando observaciones de 13 estaciones meteorológicas distribuidas en el estado de Jalisco (detalles en Cuadro 1). Las estaciones meteorológicas utilizadas pasaron por un control de calidad que consiste en que tengan datos al menos el 95% de los días en el periodo de estudio. Este procedimiento permitió ajustar los datos modelados a condiciones locales, considerando las observaciones en puntos específicos. Las estaciones utilizadas abarcan una variedad de climas y altitudes, lo que contribuye a reducir los sesgos inherentes a los productos globales.

Para integrar las observaciones puntuales de 13 estaciones meteorológicas con los datos de reanálisis ERA5 y CHIRPS en la región de estudio, las observaciones se interpolan para generar un campo espacial continuo de precipitación y temperatura utilizando el método de ponderación inversa a la distancia (Maraun, 2016; Li y Heap 2014; ). Este método se adaptó para la precipitación y temperatura interpolada en cada punto de la malla de ERA5 y CHIRPS como:

$$P_{interpolado}(x_i, y_i, t) = \frac{\sum_{i=1}^{N} w_i(x,y) \cdot P_{obs}(x_i, y_i, t)}{\sum_{i=1}^{N} w_i(x,y)}$$

Aquí, $P_{obs}(x_i, y_i, t)$ es la variable observada en la estación $i$, $w_i(x,y) = d^{-2}$ es el peso asignado inversamente proporcional al cuadrado de la distancia $d_i$ entre la estación $i$ y el punto de malla $(x,y)$, $N$ es el número total de estaciones.

Para corregir el sesgo de ERA5 y mejorar la representación de la precipitación de CHIRPS, se aplicó un ajuste multiplicativo basado en la relación promedio entre las observaciones interpoladas y los valores de ERA5 en las ubicaciones de las estaciones. El campo ajustado se calculó como:

$$P_{ajustado}(x, y, t) = P_{ERA5-CHIRPS}(x, y, t) \cdot \frac{\overline{P_{obs}(x, y, t)}}{\overline{P_{ERA5-CHIRPS}(t)}}$$

donde $\overline{P_{obs}(x,y,t)}$ y $\overline{P_{ERA5-CHIRPS}(t)}$ son los promedios espaciales de las observaciones interpoladas y de ERA5, respectivamente, calculados en las ubicaciones de las estaciones.

Finalmente, se combinaron los campos ajustados de ERA5 y CHIRPS con los campos interpolados de las estaciones para generar un campo de precipitación y temperatura mejorados ($P_{mejorado}(x,y,t)$). La combinación se realizó mediante una ponderación que considera la incertidumbre asociada a cada fuente de datos:

$$P_{mejorado}(x, y, t) = w_{obs}(x, y) P_{interpolado}(x, y, t) + [1 - w_{obs}(x, y)] P_{ajustado}(x, y, t)$$

donde $w_{obs}(x,y) = \frac{\frac{1}{\sigma^2_{ERA5-CHIRPS}}}{\frac{1}{\sigma^2_{ERA5-CHIRPS}} + \frac{1}{\sigma^2_{obs}}}$ es el peso asignado a las observaciones y $\sigma^2_{ERA5-CHIRPS}$ y $\sigma^2_{obs}$ representan las varianzas de ERA5-CHIRPS y de las observaciones, respectivamente.

## Resultados
### ENOS y otras fuentes de variabilidad

Durante la temporada de ciclones tropicales de 2024 (en el Pacífico del 15 de mayo al 30 de noviembre), se registraron ocho tormentas tropicales, un huracán de categoría 2, un huracán categoría 3 y uno de categoría 5. (Fue una temporada con menos actividad ciclónica respecto de la media histórica NHC Data In GIS Formats, 2024). Aunque ninguno de estos fenómenos impactó directamente en Jalisco, sí contribuyeron a la generación de precipitaciones (NHC NOAA, 2024). También la fase convectiva de la MJO generó condiciones propicias para la precipitación en la primera quincena de junio, a finales de julio y principios de agosto, así como en la tercera semana de septiembre y octubre (CPC - Climate Weather Linkage: Madden-Julian Oscillation, s. f.). Mientras que el ENOS presentó condiciones neutrales desde mayo hasta septiembre, lo que implicó que las precipitaciones no estuvieron significativamente influenciadas por este fenómeno.

### Características de la temporada de lluvias 2024: Anomalías de la precipitación

Con base en los datos del Servicio Meteorológico Nacional (SMN, 2024) la temporada de lluvias en Jalisco inicia a mediados de junio y concluye a finales de octubre. El tipo de precipitación característica durante la temporada de lluvias es en forma de chubascos (lluvia intensa de corta duración), así como de tormentas eléctricas; algunas de las cuales pueden llegar a tener características de severidad, como lo son rachas de viento superiores a 60 km/h, granizo de cualquier tamaño, así como la formación de tornados (León-Cruz *et al*., 2024; Díaz-Fernández *et al*., 2025).

El mes de mayo presentó un déficit en la precipitación de 10 a 30 mm con respecto a los valores climáticos, especialmente en la región Centro Oriente (Fig. 2a). El comportamiento se atribuye al inicio retrasado de la temporada ciclónica en el Pacífico mexicano, aunado a un sistema anticiclónico que se posicionó sobre el territorio nacional, mismo que permaneció hasta inicios de junio (Figura 3a), típico durante la fase El Niño.

Con el inicio de la temporada de lluvias en Jalisco, la variabilidad en las precipitaciones se hizo evidente. Junio registró valores por arriba de la media climática, salvo las regiones Costa y Sur con valores aún por debajo del promedio el resto de las regiones presentaron anomalías positivas, especialmente Centro y Norte, con anomalías de 150 – 200 mm (Fig. 2b) . El mes de julio las anomalías fueron positivas en todo el estado, salvo algunos puntos de la franja costera que se mantuvieron dentro del promedio climático, las regiones Sur y Costa Sierra presentaron anomalías de 130-160 mm.

Durante agosto y septiembre (Fig. 2d, e) las anomalías fueron positivas (100 - 150 mm) principalmente en el centro de Jalisco, mientras que las regiones Altos, Sur y Costa presentaron anomalías negativas de 80 – 130 mm. Finalmente, en octubre las anomalías de precipitación estuvieron dentro del promedio o por debajo del mismo en gran parte de Jalisco (excepto puntualmente la Costa Sur). Siendo las regiones Sur, Sierra de Amula y Costa Sierra Occidental las que presentaron las anomalías negativas más altas, 100 – 150 mm (Fig. 2f).

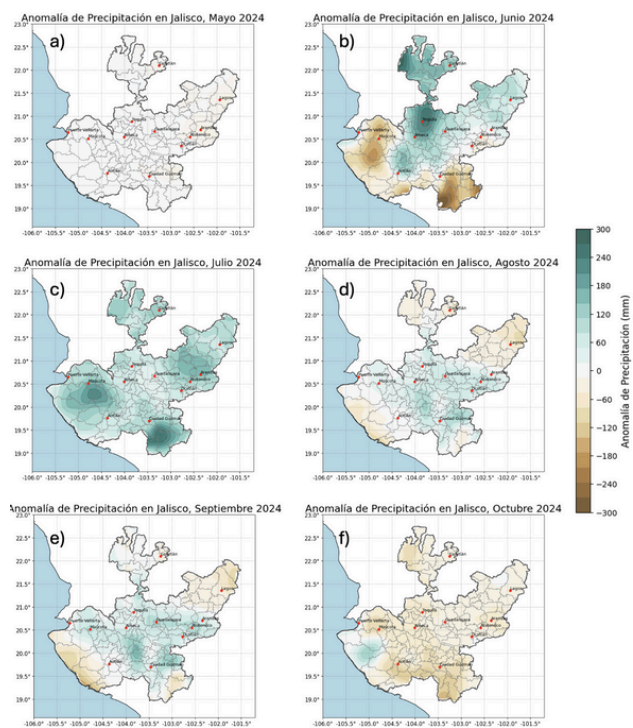

Figura 2. Anomalías de precipitación en Jalisco, para los meses de a) mayo, b) junio, c) julio, d) agosto, e) septiembre y f) octubre.
Fuente: Elaboración propia con datos de CHIRPS y estaciones meteorológicas.

En términos generales, la temporada de lluvia 2024 presentó mayor precipitación del centro al sur del estado, especialmente en regiones Sur, Costa Sur y Costa Sierra Occidental, registrando valores entre los 1200 a 1500 mm de precipitación (Fig. 3a). En balance general, Jalisco presentó valores de anomalía positiva en gran parte del estado, mientras que el déficit (alrededor de 80-200 mm) se presentaron en la franja costera y los límites con el estado de Colima y Michoacán. Con base en la Figura 3c, la estación lluviosa tiene su comienzo después del 20 de julio en todas las regiones, una vez que el sistema anticiclónico en niveles medios (Fig. 4a, b) se debilitó y permitió el ingreso de humedad y la influencia de ondas tropicales.

Las precipitaciones de las regiones de Altos Norte presentaron una disminución durante los últimos días de julio y la primera quincena de agosto (Fig. 3c), esta situación coincide con el periodo en el que la sequía de medio verano (intraestival) suele aparecer. No se muestra en este artículo, pero se puede comprobar que durante este periodo un sistema anticiclónico se fortaleció sobre el oriente y parte del centro de México, lo que favoreció la estabilidad atmosférica y por lo tanto, la disminución de las precipitaciones.

Finalmente, la temporada de lluvias presentó una disminución muy marcada a partir mediados del mes de septiembre, especialmente en Altos Norte, Norte y Costa Sur, situación que se extendió durante el mes de octubre; por esta razón, se tuvo un término anticipado de la temporada lluviosa en la mayor parte de Jalisco. No obstante, alrededor del 10 de octubre, una zona de baja presión frente a las costas de Jalisco favoreció un periodo de precipitaciones generalizadas en todo el estado, (en mayor cantidad en la costa; Fig. 3c).

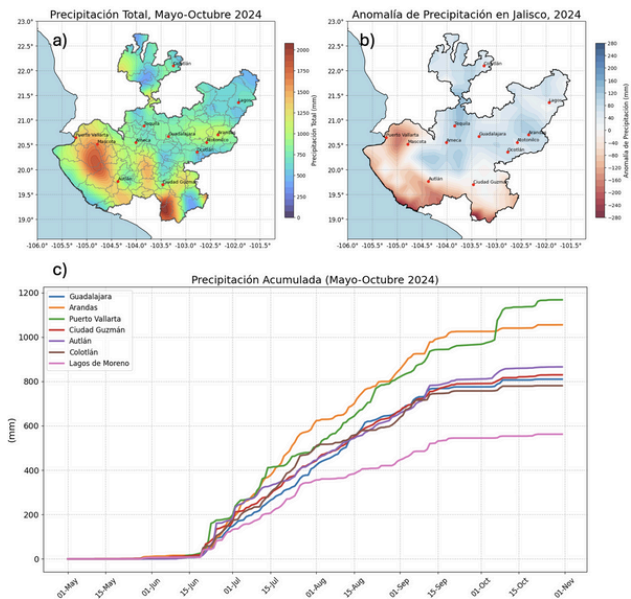

Figura 3. Comportamiento de la precipitación durante los meses de mayo a octubre en Jalisco, a) precipitación acumulada, b) anomalías de precipitación y c) lluvia acumulada en puntos específicos de las regiones de Jalisco.
Fuente: Elaboración propia con datos de CHIRPS y estaciones meteorológicas.

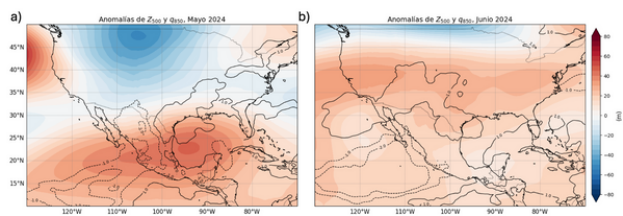

Figura 4. Anomalías de $Z_{500}$ (sombreado) y de $q_{850}$ (contornos) para a) mayo y b) junio (metros geopotenciales) de 2024.
Fuente: Elaboración propia con datos de ERA5.

**Comportamiento de las temperaturas**

Con base en los datos de las estaciones meteorológicas y ERA5, el bimestre mayo-junio es el más cálido del año, siendo mayo el más cálido. Durante el mes de mayo de 2024, se alcanzaron anomalías positivas de hasta 4°C en el oriente y Altos de Jalisco (Fig. 4a). Al interior del estado las anomalías estuvieron alrededor de 1.5 y 3°C. Mientras que en la franja costera las anomalías de temperatura fueron negativas (alrededor de 0.5 - 1.5°C). Durante este periodo, la estación meteorológica del Instituto registró una temperatura máxima de 38.4 °C el día 10 de dicho mes, muy cercano al récord histórico de 1994 de 38.6 °C. El mes de mayo de 2024 ha sido el más cálido, al menos desde 1970 en Jalisco (Fig. 5a).

El resultado del comportamiento de las temperaturas de mayo, estaría asociado a un sistema anticiclónico (alta presión en superficie) que abarca gran parte de México, mismo que se extendió hasta principios de junio (Fig. 4). Dicho sistema favorece estabilidad atmosférica, cielos despejados y en consecuencia intensa radiación en superficie. Lo que provocó un ambiente extremadamente cálido, no solo en Jalisco, sino en gran parte de México.

Junio (Fig. 5b) presentó un patrón similar al mes anterior con anomalías positivas en gran parte del estado, especialmente en las regiones Centro, Oriente y Altos de Jalisco (1.5 – 2 °C), negativas en zona costera (1 – 1.5 °C). De manera general, julio (Fig. 5c), agosto (Fig. 5d), y octubre (Fig. 5f) continuaron con anomalías de temperaturas positivas (1- 1.5°C) en todo el estado, salvo algunos puntos del centro norte y Altos de Jalisco donde las anomalías fueron negativas (0.5 – 1°C). Finalmente septiembre (Fig. 5e) presentó anomalías positivas en todo el estado del orden de 1 – 1.5 °C.

Con base en los datos de reanálisis ERA5 y los registros observacionales, la anomalía de temperatura en Jalisco durante el periodo mayo–octubre de 2024 fue de 2.3°C, superando el récord previo de 2.1°C establecido en 2023. Esto sugiere que 2024 probablemente se convertirá en el año más cálido registrado en esta región. Es importante destacar que esta anomalía de temperatura excede tanto las medias nacional como global, según el Grupo Intergubernamental de expertos sobre Cambio Climático (IPCC, 2023), el organismo científico internacional encargado de evaluar y sintetizar la investigación más reciente sobre el cambio climático y sus impactos. Esto indica que Jalisco está experimentando un calentamiento más pronunciado en comparación con estas escalas más amplias. Este aumento localizado de la temperatura podría intensificar los impactos en el sector agropecuario, particularmente en el contexto del cambio climático, donde se ha demostrado que las temperaturas extremas están asociadas con una disminución en los rendimientos agrícolas y una mayor vulnerabilidad a las sequías (Lobell *et al.*, 2011; IPCC, 2021).

En el sector agrícola de Jalisco, el aumento de las temperaturas está estrechamente relacionado con un incremento en la tasa de evapotranspiración de los cultivos. En particular, para las diferentes variedades de maíz, este aumento en la temperatura contribuye a acelerar el proceso de crecimiento, debido al incremento de los grados día de desarrollo necesarios para completar su ciclo fenológico (Lobell *et al.*, 2011). No obstante, el incremento térmico también eleva la demanda hídrica en un período de tiempo más corto, lo que, en ausencia de precipitaciones adecuadas, incrementa la probabilidad de que se registre estrés hídrico, especialmente en aquellas zonas agrícolas que dependen exclusivamente de la temporada de lluvias y carecen de sistemas de riego (Ruiz-Sinoga y Martínez-Murillo, 2009; FAO, 2016). Esta situación podría comprometer seriamente la productividad del maíz y otros cultivos para la región.

En el caso del agave azul (*Agave tequilana* Weber var. azul), el aumento en las temperaturas no sólo se traduce en máximos diurnos más elevados, sino también en un incremento de las temperaturas nocturnas. Esta condición reduce el número de horas frescas necesarias para favorecer la acumulación de azúcares en los tejidos de la planta, un proceso fundamental para la calidad del agave destinado a la producción de tequila.

Estudios previos han demostrado que temperaturas nocturnas más frescas favorecen la eficiencia en la fijación de carbono y en la acumulación de reservas energéticas en el agave (Pimienta-Barrios *et al.*, 2001; Nobel, 1994), por lo que un entorno térmico más cálido podría comprometer tanto el rendimiento como la calidad del cultivo.

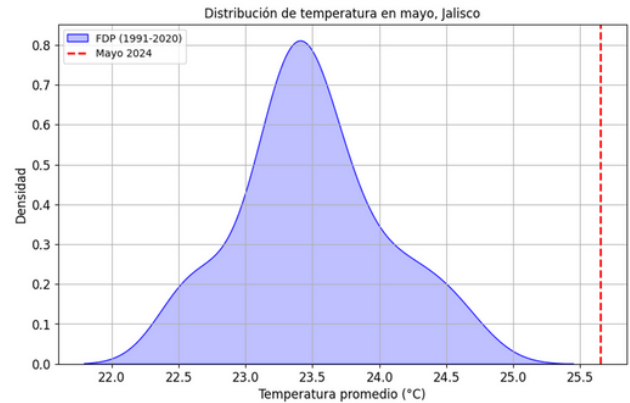

Figura 6. Distribución de la temperatura media en Jalisco durante el mes de mayo en el periodo de 1991-2020; la línea vertical roja representa la temperatura media de mayo de 2024 en Jalisco.
Fuente: Elaboración propia con datos de ERA5.

**Tormenta Eléctrica Local Severa (TELS) del 2 de julio de 2024 en el AMG**

Durante la tarde y noche del 2 y 3 de julio, se presentó un evento de TELS en el AMG. El paso de una onda tropical por el sur-occidente de México (no se muestra), así como convergencia de aire húmedo en niveles bajos y alta inestabilidad en la atmósfera favoreció el crecimiento de nubes de desarrollo vertical durante la tarde y noche del 2 de julio.

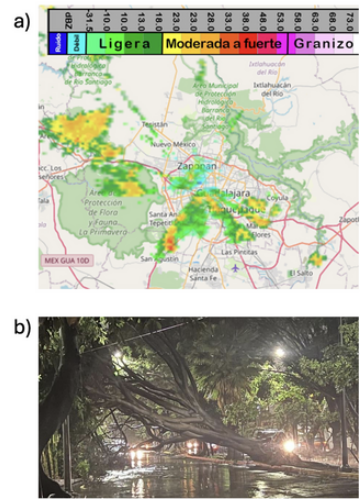

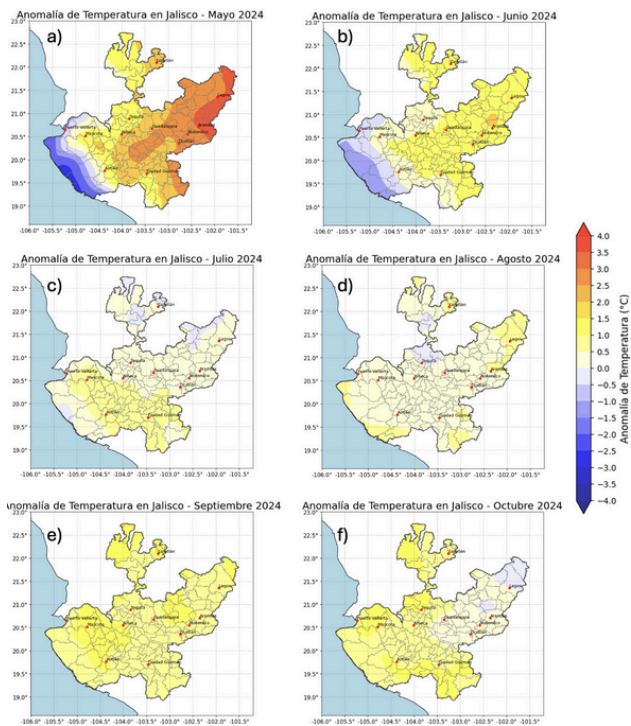

Figura 5. Anomalías de temperatura en Jalisco, para los meses de a) mayo, b) junio, c) julio, d) agosto, e) septiembre y f) octubre.
Fuente: Elaboración propia con datos de ERA5.

Figura 7. a) Imagen de reflectividad (dB) del Radar Doppler de la Universidad de Guadalajara a las 05:00 UTC del 3 de julio de 2024. Las regiones en color rojo (>40 dB) indican los núcleos con precipitación intensa. b) Daños provocados por los intensos vientos lineales asociados con la tormenta en el AMG.
Fuente: Radar Doppler de la Universidad de Guadalajara y redes sociales.

Durante la tormenta, las imágenes de reflectividad del Radar Doppler del Instituto de Astronomía y Meteorología de la Universidad de Guadalajara detectaron núcleos con reflectividad superior a 50 dB en los municipios de Zapopan, Tlajomulco y Tlaquepaque (Fig. 7a). La TELS estuvo acompañada de granizo pequeño (<2 cm) y rachas de viento superiores a 100 km/h (Cuadro 2). Algunos de los efectos de este evento fueron la caída de decenas de árboles, cableado eléctrico y anuncios espectaculares (Fig. 7b).

Cuadro 2. Racha máxima de la velocidad del viento durante el evento de la TELS en estaciones meteorológicas en el AMG.

| Nombre de la estación | Velocidad del viento | Hora (local) |
|---|---|---|
| PCYBZ_Plaza del Sol | 108.0 km/h | 21:22 h |
| GEOSMET_GDL_Sur | 102.0 km/h | 21:19 h |
| GDL_Zapopan Costco | 23.8 km/h | 18:35 h |
| AXLZAPMX | 51.5 km/h | 21:25 h |
| Estación IAM | 43.5 km/h | 21:20 h |

## Conclusiones

En este trabajo se analizaron las características de la temporada de lluvias y las temperaturas en Jalisco durante 2024, utilizando datos de reanálisis ERA5, el producto CHIRPS y 13 estaciones meteorológicas. Los resultados mostraron un inicio tardío y una conclusión anticipada de la temporada de lluvias, con anomalías positivas de precipitación en las regiones centro y sur, mientras que la franja costera presenta déficits significativos. En términos de temperatura, 2024 ha sido el año más cálido registrado en el estado, con una anomalía promedio de 2.3°C durante mayo–octubre, superando el récord previo de 2023. Este calentamiento probablemente fue impulsado por la combinación del cambio climático antropogénico y un evento de El Niño, que alteraron los patrones atmosféricos locales y afectaron la distribución de las precipitaciones.

Estos resultados evidencian la vulnerabilidad climática de Jalisco, donde las variaciones en la precipitación y la temperatura tienen impactos directos en sectores clave, particularmente en la agricultura de temporal. Las condiciones extremas de 2024—marcadas por temperaturas récord y precipitaciones irregulares—seguramente tuvieron impactos en los rendimientos agrícolas, los recursos hídricos y la dinámica urbana (especialmente por lluvias intensas y por las TELS).

La agricultura, especialmente cultivos dependientes de las lluvias de temporada como el maíz, enfrenta riesgos crecientes debido a la dependencia de patrones climáticos estables.

Es fundamental fortalecer las redes de monitoreo meteorológico, (ya que Jalisco posee solo un radar Doppler y estaciones meteorológicas limitadas) implementar modelos predictivos y priorizar políticas públicas que promuevan la resiliencia climática, con especial atención al manejo del agua y la adaptación agrícola. Además, aunque se ha estudiado de manera general el impacto del ENOS en la variabilidad climática de México, es necesario investigar a fondo la interacción entre el ENOS y sus impactos regionales, a fin de desarrollar estrategias sostenibles que respondan a eventos meteorológicos y climáticos cada vez más extremos. Este conocimiento puede beneficiar no solo a Jalisco, sino también a otras regiones con condiciones similares.